\documentclass{JHEP3}
\usepackage{graphicx,amsmath,amssymb}
\usepackage{epsfig}
\newcommand{\D}{{\cal D}}
\title{Dimensional reduction from entanglement in Minkowski space}
\author{Ram Brustein, Amos Yarom}

\abstract{ Using a quantum field theoretic setting, we present evidence for
dimensional reduction of any sub-volume of Minkowksi space. First, we show that
correlation functions of a class of operators restricted to a sub-volume of
D-dimensional Minkowski space scale as its surface area. A simple example of such
area scaling is provided by the energy fluctuations of a free massless quantum
field in its vacuum state. This is reminiscent of area scaling of entanglement
entropy but applies to quantum expectation values in a pure state, rather than to
statistical averages over a mixed state. We then show, in a specific case, that
fluctuations in the bulk have a lower-dimensional representation in terms of a
boundary theory at high temperature.}

\begin{document}

%\maketitle

\section{Introduction}

Physics up to energy scales of about one TeV is very well described
in terms of quantum field theory which uses, roughly, one quantum
mechanical degree of freedom (DOF) for each point in space. This
seems to imply that the maximal entropy $S(V)$, or the
dimensionality of phase space, of a quantum system in a sub-volume
$V$ is proportional to $V$. The Bekenstein entropy bound \cite{BEB},
that can be applied to any sub-volume $V$ in which gravity is not
dominant tells us that $S(V)$ has to be less than or approximately
equal to the boundary area $A$ of $V$ in Planck units . An
interpretation of this, relying on arguments involving physics of
black holes, was proposed by 't Hooft and by Susskind \cite{HOL} who
suggested that the number of independent quantum DOF contained in a
given spatial sub-volume $V$ is bounded by the surface area of the
region measured in Planck units. The holographic principle (see
\cite{bousso} for a recent review) postulates an extreme reduction
in the complexity of physical systems.  It is widely believed that
quantum gravity has to be formulated as a holographic theory, and it
is implicitly assumed that gravity is somehow responsible for this
massive reduction of the number of DOF. This point of view has
received strong support from the AdS/CFT duality (see \cite{adscft}
for a review), which defines quantum gravity non-perturbatively in a
certain class of space-times and explicitly exposes only the
physical variables admitted by holography.

Another route leading to area dependent entropy originated in
evidence of dimensional reduction due to quantum entanglement,
mainly from the calculation of the von Neumann entropy of
sub-systems. Srednicki \cite{srednicki} (and previously Bombelli et.
al. \cite{bombal}) considered the von Neumann entropy of quantum
fields in a state defined by taking the vacuum and tracing over the
DOF external to a spherical sub-volume of Minkowski space. They
discovered that this ``entropy of entanglement" was proportional to
the boundary area of the sub-volume.  It is possible to show that
the entropy obtained by tracing over the DOF outside of a sub-volume
of any shape is equal to the one obtained by tracing over the DOF
inside this sub-volume \cite{srednicki} but an explicit numerical
calculation was needed to show that it is linear in the area for a
spherical sub-volume. Following this line of thought, some other
geometries were considered, in particular, half of Minkowski space
in various dimensionalities (see for example,
\cite{Holzhey:1994we,callwil,sussugl,alwis1,alwis2}). Dimensional
reduction in this case seems to be quite different from that in AdS,
and its relationship to dimensional reduction due to gravity was
never clarified (see however \cite{maldacena,kleban,shenker}).

Following a calculation in \cite{befo}, we have discovered that in
the vacuum state, energy fluctuations of a free massless scalar
field in a sub-volume of Minkowski space are proportional to the
boundary area of this sub-volume.  Energy fluctuations of quantum
fields in their vacuum state in the whole of Minkowski space vanish,
of course, because the vacuum is an eigenstate of the hamiltonian.
But, energy fluctuations in a sub-volume of Minkowski space do not
vanish \footnote{
    As the sub-volume in question is of the order of the volume of
    Minkowski space, the energy fluctuations rapidly decrease to
    zero.}. Not only do they not vanish, but they are divergent. Once a
high-momentum cutoff is introduced to regularize the divergence,
they depend on a positive power of the cutoff. The power is
determined by dimensional analysis such that the result has the
correct dimensionality. We will discuss the significance of the
divergences in detail after presenting a concrete example.

The origin of such energy fluctuations is, in some sense, similar to the origin of
entanglement entropy discussed above except that they do not involve tracing over
DOF, and are calculated as quantum expectation values in a pure state, rather than
as statistical averages over a mixed state. Area scaling is not specific to energy
fluctuations of free massless fields. We show that it is valid for correlation
functions of a class of operators and persists for interacting field theories.
Here, also, the correlation functions diverge, with a power dependence on the UV
cutoff. The power being determined by the dimensionality of the operator. In a
specific case we show that the area scaling of correlation functions results from a
boundary field theory at high temperature.

Let us start with a simple example: energy fluctuations  of a free
massless quantum field theory in its vacuum state in a sub-volume
$V$ of Minkowski space. We define the (normal-ordered) energy
operator for this sub-volume $ E^V=\int_V :\!\mathcal{H}(\vec{x})\!:
d^dx $, where $:\mathcal{H}(\vec{x}):$ is the normal-ordered
zero-zero component of the energy-stress tensor, i.e., the
hamiltonian density. For a free massless field theory in $d$ spatial
dimensions it is possible to express $E^V$ in terms of canonical
creation and annihilation operators $a^\dagger_{\vec{p}}$ and
$a_{\vec{p}}$: $
    E^V = \frac{1}{4}\frac{1}{(2\pi)^{2d}}
        \int_V d^dx\int
        \frac{d^dp}{\sqrt{p}}\,\frac{d^dq}{\sqrt{q}}
        e^{\imath(\vec{p}+\vec{q})\cdot \vec{x}}$
       $ \left\{2\left(p q -\vec{p} \cdot \vec{q} \right)
            a^\dagger_{\vec{p}}a_{\vec{q}} -
            \left(\vec{p} \cdot \vec{q} + p q \right)
            \left(a_{\vec{p}}a_{\vec{q}} +
                a^\dagger_{\vec{p}}a^\dagger_{\vec{q}} \right)
        \right\}.
$ The vacuum expectation value of $E^V$ vanishes: $\langle 0| E^V
|0\rangle=0$. However, the energy fluctuations do not:
\begin{equation}
\label{E:HV2}
    \langle 0|(E^V)^2|0\rangle = \frac{1}{8} \frac{1}{(2\pi)^{2d}}
               \int_V d^dy_1\int_V d^dy_2 \int d^dp \int d^dq
                e^{\imath(\vec{p}+\vec{q})\cdot
               (\vec{y}_1-\vec{y}_2)}
               {\left(\frac{\vec{p} \cdot \vec{q}}
                    {\sqrt{ p  q}} +
                    \sqrt{p q}\right)}^2.
\end{equation}
After performing the integration over the momenta, the integrand is a function of
the variable $|\vec{y}_1\!-\!\vec{y}_2|$. Rewriting eq.({\ref{E:HV2}) as
$\langle(E^V)^2\rangle= \int \limits_0^\infty F(y) \D_V(y) dy, $ where
$$
    F(y)=\frac{1}{8} \frac{1}{(2\pi)^{2d}}
           \int
            \left( pq+
                2\vec{p} \cdot \vec{q}+
                \frac{(\vec{p} \cdot {\vec{q}})^2}{pq} \right)
            e^{-\imath(\vec{p}+\vec{q})\cdot \vec{y}}
                d^dp\,d^dq,
$$ and
\begin{equation}
\label{hol5}
    \D_V(y)=\int_V d^dy_1\int_V d^dy_2 \,
            \delta^{(d)}(y-|\vec{y}_1-\vec{y}_2|),
\end{equation}
the integral is separated into a geometric factor, $\D_V(y)$,
which depends only on the shape of the sub-volume $V$, and another
factor $F(y)$.

Evaluating $\langle 0|(E^V)^2|0\rangle$ we find, remarkably, that
it is proportional to the boundary area $S(V)$ of $V$ for any
volume whose boundary is regular. It is possible to compute the
energy fluctuations analytically with an exponential  high
momentum (UV) cutoff $\Lambda$:
$\langle 0|(E^V)^2|0\rangle
    =\frac{
        \Gamma\left(\frac{d+1}{2}\right)
        \Gamma\left(\frac{d+3}{2}\right)}
        {\Gamma\left(2+\frac{d}{2}\right)}
    \frac{1}{2^{d+4}\pi^{\frac{d}{2}+1}} \Lambda^{d+1} S(V)$.
(See \cite{bruyar} for details.) These fluctuations can also be
evaluated numerically using different cutoff schemes \cite{BYO}. One
can prove that the area dependence is a robust result, while the
numerical coefficients depend on the details of the cutoff
procedure. This area scaling behavior may have been inferred by
noting that if we denote the complement of $V$ by $\widehat{V}$ and
that $S(V)=S(\widehat{V})$ then
    $\langle 0|(E^{\widehat{V}})^2|0\rangle=\langle0|(E^V)^2|0\rangle$.

The surprising result that energy fluctuations in a sub-volume of
Minkowski space are proportional to the boundary area of the
sub-volume is not accidental. Rather, it seems to be a general
characteristic property of sub-systems. We wish to understand this
fact, and examine how general it is, which we do in the rest of
the paper.

\section{Operators in a finite sub-volume and their correlation functions  }

As a generalization of the above, we wish to examine expectation values of
operators of the form $\langle O^V_i O^V_j \rangle$ defined as follows. Let us
consider a  finite $d$ dimensional space-like domain of volume $V$ and linear size
$R$, which we think  of as being part of a $d=D-1$ dimensional Minkowski space. A
quantum field theory is defined in the whole space. Generically, it is an
interacting field theory, which comes equipped with a high momentum (UV) cutoff
$\Lambda$, and some regularization procedure.  A low momentum (IR) cutoff is
implicitly assumed, but we will not discuss any of its detailed properties. The
sub-volume is ``macroscopic" in the sense that $R\Lambda\gg 1$. The boundary of $V$
is an ``imaginary" boundary, since we do not impose any boundary conditions, or
restrictions on the fields on it.

We will be interested in a set of  (possibly composite) operators
$O_i$ which can be expressed as an integral over a density ${\cal
O}_i$: $ O_i=\int d^d x {\cal O}_i(\vec{x})$. For this set we may
define the operators $ O_i^V=\int_V d^d x {\cal O}_i(\vec{x}) $. We
will discuss for concreteness the case in which the large space is
in the vacuum state $|0\rangle$. Our results can be easily
generalized to states other than the vacuum. We note that the vacuum
will not be an eigenstate of $O_i^V$.

We have already considered energy fluctuations in a given sub-volume
$V$. In general, in addition to fluctuations of operators, namely,
two point functions of the same operator, we may look at two point
functions of different operators. Such correlation functions also
scale linearly with the boundary area of the sub-volume $V$. To show
this, we would like to evaluate the correlation function $\langle
0|O_i^{V}O_j^{V}|0\rangle= \int\limits_V d^d y_1 \int\limits_{V} d^d
y_2~ F_{ij}\left(|\vec{y}_1-\vec{y}_2|\right)$. Following
\cite{befo}, we evaluate all the integrals except for the
$y=|\vec{y}_1-\vec{y}_2|$ integral, $\int\limits_V d^d y_1
\int\limits_{V} d^d y_2~
F_{ij}\left(|\vec{y}_1-\vec{y}_2|\right)=\int\limits_0^{\infty} dy~
\D_V(y)~ F_{ij}(y)$.

The geometric factor $\D_V(y)$ can be evaluated explicitly. By a careful geometric
analysis, one can show that $D_V(y)=G_V V y^{d-1}+G_S S(V)
y^d+\mathcal{O}(y^{d+1})$, $G_V$ and $G_S$ being constants, and $S(V)$ the surface
area of $V$ \cite{bruyar}. We will need to use $ \D_V(y)\sim y^{d-1}$ for $y\sim
0$. This property may be shown by considering eq.(\ref{hol5}). The limit $y\sim 0$
is approached when both vectors  $\vec{y}_1$ and $\vec{y}_2$ are almost equal.
Fixing one of them, say $\vec{y}_1$, at some arbitrary value makes the integral
over $\vec{y}_2$ an unrestricted $d$-dimensional integral, which in spherical
coordinates yields the above result. Defining ${\cal D}_V(y) = y^{d-1}R^d\widetilde
\D_V\left(\frac{y}{R}\right)$, which captures the function's behavior near $y\sim
0$, and its dimensionality (recall that $R$ is the linear scale of $V$). We note
that for a generic sub-volume $ {\widetilde \D_V}^{\prime}(0) \ne 0$. We would like
to emphasize that in general $R^d$ will be replaced by a volume factor and
$R^{d-1}$ will be replace by the surface area of the boundary as discussed
exhaustively in \cite{bruyar}. We will continue the analysis using the expressions
$R^d$, and $R^{d-1}$ as the general case is somewhat encumbering.

The short distance/large momentum  behavior of  correlation
functions is determined by the full mass dimensionalities
$\delta_i$, $\delta_j$ (including anomalous dimensions induced by
interactions)  of the operator densities ${\cal O}_i$ and ${\cal
O}_j$.  At short distances $F_{ij}\left(|\vec{x}|\right)\sim
\frac{1}{|\vec{x}|^{\delta_i+\delta_j} }$, equivalently for large
momentum, $ \widetilde F_{ij}\left(|\vec{q}|\right) \sim
|\vec{q}|^{\delta_i+\delta_j-d}$, where we have defined
$\widetilde F_{ij}\left(|\vec{q}|\right)= \int ~ d^dx
 ~ e^{i\vec{q}\cdot\vec{x}}
F_{ij}\left(|\vec{x}|\right)$. Here we have used the leading
short-distance behavior.  In general, $\widetilde
F_{ij}\left(|\vec{q}|\right)$ will be a function of several powers
of $q$, which may be fractional or perhaps multiplied by logarithms.
We will consider operators whose correlation functions are sharply
peaked at short distances, and decay at large distances. The short
distance/large momentum behavior of correlation functions is in many
cases singular before a UV cutoff is introduced and needs to be
regularized. We do this by cutting off the momentum integrals that
appear in the expressions for the two-point functions. The
significance of these divergences will be discussed in detail later.
The details of the regularization procedure and the exact nature of
the UV cutoff are not particularly important since our main
observation is the area scaling nature of the fluctuations.

So after all is said and done, we need to evaluate the following
integral,
\begin{equation}
 \label{corr3}
   \int\limits_0^{\infty} dy~ {\cal D}_V(y)~ F_{ij}(y)    =
   R^d\int\limits_0^{\infty} dy~ y^{d-1}
  ~\widetilde{\cal D}_V\left(\frac{y}{R}\right)
    \int \hbox{\large$\frac{d^dq}{(2\pi)^d}$} ~ e^{-iqy \cos\theta_q}
        \widetilde{F}_{ij}(|\vec{q}|),
\end{equation}
where $\theta_q$ is the angle of the vector $\vec q$ with respect to
some fixed axis. The last momentum integral is regularized by the UV
cutoff of the theory. Without such a cutoff it is divergent in the
large momentum region whenever $\delta_i+\delta_j>0$. We choose to
regulate the integral in eq.(\ref{corr3}) by an additional factor
that suppresses the high momentum contribution.

Integrals of the form appearing in eq.(\ref{corr3}) can be evaluated
analytically for $F(|\vec{q}|)\propto(q^2)^n$ with $n$ integer and a
gaussian momentum, and for $F(|\vec{q}|)\propto q^\alpha$ with
$\alpha$ rational and an exponential momentum cutoff. We will
discuss in some detail the integer case, and quote the results for
the other case.

For a Gaussian cutoff eq. (\ref{corr3}) with $F_{ij}(q)\sim
q^{2n}$, reduces to
\begin{equation}
 \label{eyeen1}
I_n=R^d\int\limits_0^{\infty} dy~ y^{d-1}~\widetilde{\cal
D}_V\left(\frac{y}{R}\right) \int d^dk ~ e^{ i k y
\cos\theta_k-\frac{1}{2} k^2/\Lambda^2} (k^2)^n.
\end{equation}
Naively, this integral scales as $R^d\Lambda^{2 n}$ (extensively)
as the pre-factor might suggest. Now, $|\vec{k}|^2$ may be
rewritten as a d-dimensional laplacian in spherical coordinates:
$\frac{1}{y^{d-1}}\partial_y y^{d-1}
\partial_y$. If we now integrate by parts once, the volume term will vanish. Due to the
gaussian, we may approximate $y\sim 0$, and we are left with
\begin{equation}
\label{eyeen2} I_n=a_V~C_d~R^{d-1} \Lambda^{2n-1},
\end{equation}
where $C_d\sim (-1)^n\int\limits_0^{\infty} dx
        x^{d-2} \left[
        \left(
            \frac{1}{x^{d-1}}\partial_x x^{d-1} \partial_x
        \right)^{n-1}
        \hbox{\large $e$}^{ \hbox{${-\frac{1}{2} x^2}$}}
        \right]$
is a finite $d$- and $n$-dependent remnant of the momentum
integral, and $a_V= (2\pi)^{d/2}~ \widetilde{\cal D}'_V(0)$. Thus
we have shown that $I_n$ scales as $R^{d-1}\Lambda^{2n-1}$.
Similarly $ I_\alpha=R^{d} \int\limits_0^{\infty} dy~y^{d-1}
~\widetilde{\cal D}_V\left(\frac{y}{R}\right) \int d^dk ~ e^{
i\vec{k}\cdot\vec{y}-k/\Lambda} k^\alpha\sim \Lambda^{\alpha-1}
R^{d-1} $, where $\alpha$ is not necessarily integer.

It is possible to show that the area scaling law $\langle
O_{i}^VO_{j}^V \rangle \propto
 \Lambda^{\delta_i+\delta_j-1}R^{d-1}$ is robust to changes in the
cutoff procedure. Any cutoff function $C(k/\Lambda)$ will give
similar results provided that it and all of its derivatives vanish
at $k\rightarrow\infty$, and that its integral $\int d^d k
C(k/\Lambda)$ is finite. It is also possible to show \cite{bruyar}
that if the connected two point function of the operator densities
$F_{ij}(y) \equiv \nabla^2 g_{ij}(y) $ satisfies $(i)\ g_{ij}(y)$ is
short range: at large distances it decays fast enough
$|g^{\prime}(y)| < 1/y^{a+1}$, with $a \geq d-1$, and $(ii)\
g_{ij}(y)$ is not too singular at short distances: for small $y$,
$|g(y)| < 1/y^{a+1}$ with $a<d-2$, then the connected two point
function $ \langle O^{V_1}_i O^{V_2}_j \rangle_C $ is proportional
to the area of the common boundary of the two regions $V_1$ and
$V_2$. We would like to emphasize that our results are valid for any
function that obeys the above conditions, and not only by
correlation functions that have strictly power law dependence on the
distance.

The divergence of the correlation functions cannot be renormalized in the standard
way by adding local counter terms to the action. This is because the coefficient of
the divergence depends on the area, and therefore on the geometry of the imaginary
sub-volume in which they are calculated. Additionally, they appear at the level of
free field theory which is usually not renormalized. This suggests that the
divergence of the correlation functions is related to a physical UV cutoff of the
theory given by, say, a Planck scale lattice or some other UV complete theory. If a
physical UV completion of the field theory exists, then in the framework of that UV
completion the divergences become indeed finite and the cutoff has a physical
meaning.

\section{Lower-dimensional representation of correlation functions}
The fact that correlation functions of operators scale as the
boundary area of the sub-volume rather than as the volume (as
expected), suggests that it might be possible to find a
representation of them on the boundary $\partial V$ of $V$ (and of
$\widehat {V}$).  To show this we shall express $\langle
0|O_i^{V}O_j^{V}|0\rangle$ as a double derivative
$\langle 0|O_i^{V}O_j^{V}|0\rangle =
        -\int\limits_V d^d y_1
        \int\limits_{V} d^d y_2 \vec{\nabla}_{1}\cdot \vec{\nabla}_{2}
        g_{ij}(|\vec{x}-\vec{y}|) $
and use Gauss' law to express it as a boundary correlation function,
$\langle 0|O_i^{V}O_j^{V}|0\rangle =
    \oint\limits_{\partial V} d^{d-1} z_1
    \oint\limits_{\partial V} d^{d-1} z_2
    \hat{n}_1 \cdot \hat{n}_2 g_{ij}(|\vec{z}_1-\vec{z}_2|)$.
This allows us to define an equivalent correlation function in
$d-1$ dimensions of some operators $\Theta_i^{\partial V}$ and
$\Theta_j^{\partial V}$
\begin{equation}
\label{boundary4}
    \langle 0|O_i^{V}O_j^{V}|0\rangle \sim
    \langle 0|\Theta_i^{\partial V}\Theta_j^{\partial V}|0\rangle_{d-1}=
    \alpha^d_{ij} \int\limits_{\partial V}\!\! d^{d-1} z_1
    \int\limits_{\partial V}\!\! d^{d-1} z_2~\hat n_1 \cdot\hat n_2
        \langle 0|\hbox{\large$\vartheta$}_i(\vec{z}_1)
        \hbox{\large$\vartheta$}_j(\vec{z}_2)|0\rangle_{d-1}.
\end{equation}
Here $\hat n$ is a boundary unit normal. For example,  in the case
that the correlation functions are pure powers
$\frac{c_{ij}^d}{|\vec{y}_1-\vec{y}_2|^{\delta_i+\delta_j}}$, then
$\alpha^d_{ij}=\frac{c_{ij}^d}{(\delta_i+\delta_j-2)(d-\delta_i-\delta_j)c_{ij}^{d-1}}$.
In the special case $\delta_i+\delta_j=2$ the power becomes a
logarithm, and for $\delta_i+\delta_j=d$ it becomes a delta
function, so they need to be handled with special care. Note that
 $|\vec{z}_1-\vec{z}_2|$, even though evaluated for
$\vec{z}_1$ and $\vec{z}_2$ on the boundary, generally expresses
distances in the bulk. In the case that the boundary is a plane,
$|\vec{z}_1-\vec{z}_2|$ indeed expresses distances on the
boundary.

It will be very interesting to determine the consistency
conditions under which a boundary representation of integrals of
bulk correlation functions can be derived from an effective
action, and to determine how symmetries of the sub-volume and of
the bulk theory are reflected in the properties of their boundary
representations. A covariant formulation, starting from a
Lagrangian in $D=d+1$ space-time dimensions would be very useful.
At this point we would like to present evidence of a special case
where the boundary theory is a theory at high-temperature, and
leave the answers to the questions above open for future research.

We consider a massless, free scalar field theory and take $V$ to be
half of Minkowski space (under some assumptions it is possible to
generalize this to other geometries.) We first show that the n-point
(single field) functions for a (d+1) dimensional free scalar field
theory in the vacuum state are equal to n-point functions of a
((d-1)+1) dimensional free scalar field theory at high-temperature,
\begin{equation}
\label{E:dimredexp}
    \langle 0| e^{-J \int_V \phi (\vec{x}) d^dx}|0 \rangle_{d}
    = \langle e^{-J \sqrt{\alpha^d_{\phi\phi}}\int_{\partial V} \phi (\vec{x})
    d^{d-1}x}\rangle_{d-1}^{\beta \to 0}.
\end{equation}
We shall show this equality term by term, $
    \int_V d^dx_1\ldots\int_V d^dx_{2n}\langle 0|\phi (\vec{x}_1) \cdots
    \phi (\vec{x}_{2n})|0 \rangle_{d}  =
    ({\alpha^d_{\phi\phi}})^n \int_{\partial V} d^{d-1}x_1 \cdots\int_{\partial V}
    d^{d-1}x_{2n}
        \langle\phi (\vec{x}_1) \cdots \phi (\vec{x}_{2n})
        \rangle_{d-1}^{\beta \to 0}.
$ For an odd number of fields this is trivially satisfied, since
both sides vanish. For an even number of free fields in any
dimension, $
    \langle 0|\phi (\vec{x}_1) \cdots
    \phi (\vec{x}_{2n})|0 \rangle=
     \sum\limits_{\substack{\text{\tiny{all}} \\ \text{\tiny{perm.}}}}
        \langle 0|\phi(\vec{x}_{i_1})\phi(\vec{x}_{i_2})|0 \rangle \cdots
        \langle 0|\phi(\vec{x}_{i_{2n-1}})\phi(\vec{x}_{i_{2n}})|0 \rangle$.
So to prove eq.(\ref{E:dimredexp}) we only need to show that
 $$ \int_V d^dx_1\int_V d^dx_2 \langle 0|\phi (\vec{x}_1) \phi
(\vec{x}_2)|0 \rangle_{d} =
    \alpha^d_{\phi\phi} \int_{\partial V} d^{d-1}x_1\int_{\partial V} d^{d-1}x_2
        \langle \phi (\vec{x}_1) \phi (\vec{x}_2)\rangle_{d-1}^{\beta \to 0}.$$
The mass dimension of a scalar field in $d+1$ dimensions is
$\delta=\frac{d-1}{2}$, so
\begin{align}
\notag
    \int_V d^dx_1\int_V d^dx_2 \langle
        \phi (\vec{x}_1)
        \phi (\vec{x}_2) \rangle_{d} & =
    \alpha^d_{\phi\phi} \int_{\partial V} d^{d-1}x_{1_\bot}
    \int_{\partial V}d^{d-1}x_{2_\bot}
        \frac{\hat{n}_1 \cdot
        \hat{n}_2}{|\vec{x}_1-\vec{x}_2|^{d-3}} \\ & =
\label{boundaryp}
      \alpha^d_{\phi\phi}
        \int_{\partial V} d^{d-1}x_{1_\bot}
        \int_{\partial V} d^{d-1}x_{2_\bot}
        \langle \phi(\vec{x}_1) \phi(\vec{x}_2)
        \rangle_{d-1}^{\beta \to 0}.
\end{align}
In eq.(\ref{boundaryp}) we have used the fact that a $D=d+1$
dimensional field theory in the limit of high temperature is
equivalent to a $D-1$ dimensional field theory at zero
temperature, and that $\hat{n}_1\cdot\hat{n}_2=1$ when $V$ is half
of Minkowski space. Our choice of viewing the power of the
integrand as that of a $(d-1)$ dimensional theory is motivated by
the region of integration. One should note that this procedure is
quite distinct from the well-known dimensional reduction of a
high-temperature quantum field theory. Here we relate a $d+1$
dimensional theory to a $d-1$ dimensional theory, which is
equivalent to a $d$ dimensional theory at high temperature.

Similarly, such dimensional reduction may also be explicitly shown for other two
point functions
\begin{align*}
    & \int\limits_ V  d^dx_1 \int\limits_V d^dx_2 \langle 0|
    \nabla_1^m \phi^{n} (\vec{x}_1)
          \nabla_2^{m^{\prime}} \phi^{n^{\prime}}
       (\vec{x}_2) |0 \rangle_{d}
    \cong \\
   & \int\limits_{\partial V} d^{d-1}x_{1_\bot}
            \int\limits_{\partial V} d^{d-1}x_{2_\bot}
            \langle\nabla_{1}^{m+n-1}  \phi^{n} (\vec{x}_1)
            \nabla_2^{m^{\prime}+n-1} \phi^{n^{\prime}}
     (\vec{x}_2) \rangle_{d-1}^{\beta \to 0},
\end{align*} where by $\nabla_{1}^{m} \nabla_{2}^{m^{\prime}}$, we mean the scalar operator
obtained by consecutive operations of $\nabla$; in order for this to be a scalar
operator we must have that $m+m^{\prime}$ is even. This may also be generalized to
the case of single field n-point functions with an arbitrary number of derivatives.

We are able to extend our conclusions to the most general
correlation functions in the theory - products of powers of fields
and their derivatives, only in the case of a large number $N$ of
fields (in which case the correlation functions reduce to products
of two-point functions.) To see this explicitly and to discuss the
large $N$ limit we define
$\Phi(x_i)=\hbox{diag}(\phi_1(x_i),\ldots,\phi_N(x_i))$. In this
limit only correlations functions of the form $\int_V \ldots \int_V
<\hbox{Tr}\Phi(x_1)\ldots\hbox{Tr}\Phi(x_n)> d^dx_1\ldots d^dx_n$
contribute to leading (zero) order in the $1/N$ expansion. Thus, to
first order in the large $N$ limit, all bulk correlation functions
have a boundary description.

The fact that the boundary theory seems to be in a high temperature state is
reminiscent of the thermal atmosphere near black hole horizons. The local Hawking
temperature increases from its asymptotic value as one gets near the horizon. In
fact, there were some attempts to explain  black hole entropy and its area scaling
as originating from the thermal entropy of the near-horizon thermal atmosphere
\cite{Zurek:1985gd,'tHooft:1984re,Frolov:1998vs}. Another connection with black
hole physics is the area dependence of thermodynamic quantities. The fluctuations
we have described are quantum. However, if instead of quantum expectation values of
operators restricted to $V$, we consider fluctuations in the state
$\rho_V=\text{Tr}_{\widehat{V}}|0\rangle\langle0|$, we get from $\langle 0| O^V
|0\rangle=\text{Tr}(\rho_V O^V)$ that area scaling of quantum fluctuations leads to
area dependent thermodynamics \cite{bruyartherm}.

The nature of the dimensional reduction that we have discussed
seems to be more restricted than the holographic correspondence in
AdS/CFT: we have presented evidence that a class of operators that
can be expressed as an integral over a density in the bulk can be
represented by an integral over a density on the boundary. On the
other hand the AdS/CFT duality conjectures a correspondence
between local operators in the bulk and in the boundary. Since we
have not used gravity to establish our correspondence, this raises
the possibility that both gravity and quantum mechanics have
distinct functionality in holography.

\acknowledgments

Research supported in part by an Israel Science Foundation grant
174/00-2. A.Y. is supported in part by the Kreitman Foundation. We
thank G. Veneziano and D. Oaknin for extensive and fruitful
discussions and O. Aharony, M. Einhorn,  B. Kol, D. Marolf, E.
Novak, S. Zhang and M. Srednicki for comments and discussions.

\end{document}